# Unlocking High-Throughput Heterojunction Discovery

*Thomas W. Gries[1,2,§], Davide Regaldo[3,§], Yanyan Duan[1], Florian Scheler[1], Maxim Simmonds[1], Valerio Stacchini[1], Annamaria Petrozza[3], Eva Unger[1,4], Antonio Abate[1,2,5], Jean-Paul Kleider[6,7,8], Artem Musiienko\*,[9]*

[1]Helmholtz-Zentrum Berlin für Materialien und Energie (HZB), 14109, Berlin, Germany.

[2]Faculty of Chemistry, University of Bielefeld, 33615, Bielefeld, Germany.

[3]Center for Nano Science and Technology, Instituto Italiano di Tecnologia, 20134, Milan, Italy.

[4]Department of Chemistry & IRIS Adlershof, Humboldt University, 12489, Berlin, Germany.

[5]Department of Chemical, Materials and Production Engineering, University of Naples Federico II, 80125, Naples, Italy.

[6]Institut Photovoltaïque d'Île-de-France (IPVF), 91120, Palaiseau, France.

[7]Université Paris-Saclay, CentraleSupélec, CNRS, Laboratoire de Génie Electrique et Electronique de Paris, 91192, Gif-sur-Yvette, France.

[8]Sorbonne Université, CNRS, Laboratoire de Génie Electrique et Electronique de Paris, 75005, Paris, France.

[9]Young Investigator Group, Robotized Material and Photovoltaic Engineering, Helmholtz-Zentrum Berlin für Materialien und Energie (HZB), 12489, Berlin, Germany.

[§] These authors contributed equally to this work.

\*Corresponding Author: artem.musiienko@helmholtz-berlin.de

# Abstract

Photoluminescence (PL) is a ubiquitous proxy for material quality in optoelectronic devices, widely used for high-throughput materials discovery. However, we demonstrate that in the presence of charge-selective contacts, PL loses its predictive reliability and can exhibit strong quenching even in highly efficient photovoltaic devices under open-circuit conditions. By combining steady-state and transient PL with contactless transient surface photovoltage measurements we disentangle the intertwined processes of extraction and recombination, clarifying the physical origin of this phenomenon. This joint approach reveals extraction dynamics not captured by PL alone. A digital replica of the interface shows that Coulomb attraction and interfacial recombination are the fundamental mechanisms driving quenching after charge extraction. Based on these insights, we present a decision tree for heterojunction classification and PL interpretation applicable across diverse optoelectronic systems, including photovoltaics, photodetectors, and LEDs. Our approach supports systematic screening and optimization of half-devices, bridging the gap between accelerated materials discovery and accelerated device discovery.

# Introduction

Photoluminescence (PL) spectroscopy is an essential tool in semiconductor device research for characterizing optoelectronic materials and interfaces. In semiconductors, PL occurs when absorbed light generates electron-hole pairs that recombine radiatively, emitting photons. Both steady-state PL (ssPL) and transient PL (trPL) are widely used to evaluate material quality in applications such as photovoltaics,[1-3] light-emitting diodes,[4-6] lasers,[7-11] sensors,[12] and transistors.[13-15] Most recently, PL-based methods have supported automated material assessment,[16] discovery,[17,18] and optimization,[19,20] as demonstrated in studies by Brabec et al.,[17-19] Tarín et al.,[16] and Tei et al.[20] While interface quality can also be assessed under short-circuit conditions in complete devices, as demonstrated in voltage-dependent PL studies,[21] this approach requires both electrical contacts. This constraint limits the use of PL for high-throughput interface screening, where rapid, modular evaluation of individual layers is needed. Consequently, the lack of a contact-independent method for interface assessment remains a significant obstacle to accelerated interface discovery.

Despite the ubiquitous use of PL spectroscopy, with ~100,000 publications annually (**Fig. 1A**), PL measurements often lead to conflicting interpretations in the presence of CSCs. In pristine semiconductors, a bright PL (**Fig. 1B**) typically signifies high material quality with low defect densities.[22-24] However, the presence of CSCs introduces additional processes, including interface recombination and charge extraction, which complicate PL interpretation in both partial and complete device stacks.[25-29] Interface modifications such as doping,[30] passivation,[31] or insertion of dipole layers[32] influence the PL intensity, which may decrease (*quenching*) or increase (*enhancement*; **Fig. 1C**).

To support high-throughput characterization of heterojunctions and other partial device stacks, we investigate the fundamental origin of PL quenching. We complemented PL with transient surface photovoltage (trSPV, **Fig. 1D**), a contactless probe of the dynamics of charge separation. The initial slope of the trSPV signal reflects the charge extraction rate, while the SPV amplitude is proportional to the product of charge separation distance $\Delta x$ and the amount of extracted charge $q_{\text{extr}}$ ($V_{\text{SPV}} \propto \Delta x \times q_{\text{extr}}$).[33,34] We demonstrate that PL quenching under open-circuit conditions can arise from efficient charge extraction from the semiconductor to the CSC, a scenario often misinterpreted as evidence of poor interface quality. To resolve this ambiguity, we integrated PL and trSPV with a predictive drift-diffusion (DD) model that resolves the redistribution of charge carriers across the interface. This digital replica identifies the physical origin of PL quenching and supports the development of a practical decision tree for heterojunction classification and PL interpretation. Our approach enables non-invasive assessment of interface quality and accelerated materials screening without full device fabrication.

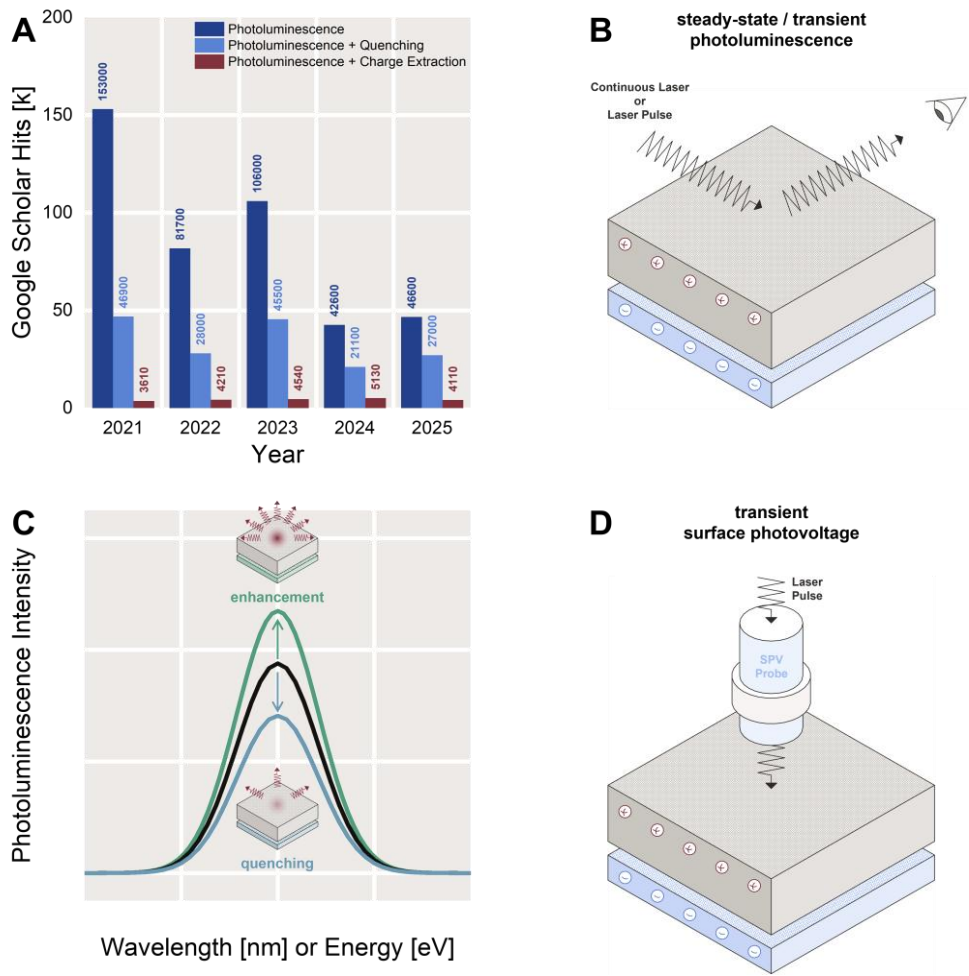

Fig. 1. **Overview of photoluminescence (PL) and transient surface photovoltage (trSPV). (A)** Annual publication volume from 2021 to 2025 involving the keywords "photoluminescence" (dark blue), "photoluminescence" and "quenching" (light blue), and "photoluminescence" and "charge extraction" (dark red), based on hits in Google Scholar. **(B)** Schematic of steady-state PL (ssPL) or transient PL (trPL) probing radiative recombination in the semiconductor absorber. trPL resolves the temporal decay of photogenerated charge carriers following pulsed excitation. **(C)** Example ssPL spectra under open-circuit conditions for semiconductor/charge-selective contact heterojunctions. Samples show either PL quenching (light blue) or PL enhancement (light green) relative to a control sample (black), depending on the interface properties. **(D)** Schematic of the principle of trSPV: Contactless detection of spatially separated charge carriers after pulsed excitation. The signal provides time-resolved information on charge extraction.

## Results and Discussion

To understand why incomplete devices show quenched PL at open-circuit conditions in various scenarios, we systematically studied three different classes of semiconductor/CSC heterojunctions commonly encountered in current state-of-the-art thin film technology, each affected by a distinct fundamental interface modification. These include (i) *defect-limited interfaces* where recombination is reduced by chemical passivation,[35] (ii) *extraction-limited interfaces* where charge extraction is hindered by energetic alignment or transport barriers,[36] and (iii) *mixed interfaces* where both defect- and extraction-limitations coexist. Each interface was characterized by using steady-state and transient PL (ssPL and trPL), photovoltaic device performance, and surface photovoltage (trSPV), as shown in **Fig. 2**.

We first investigated a heterojunction of perovskite [(FA$_{0.83}$MA$_{0.17}$)$_{0.95}$Cs$_{0.05}$Pb(Br$_{0.17}$I$_{0.83}$)$_3$, 3Cat] and nickel oxide (NiO$_x$) modified by a self-assembled monolayer (SAM), known to reduce interface recombination.[30,34,37-40] Upon SAM insertion, we observed pronounced PL enhancement (**Fig. 2A**, **Fig. S1A**) and a rise in quasi-Fermi level splitting (QFLS) from 1.10 eV to 1.18 eV (**Fig. S1B**). The trPL signal decayed more slowly (**Fig. 2B**), reflected in longer differential charge carrier lifetimes (**Fig. S1C**). PL enhancement and longer charge carrier lifetimes indicate reduced interface losses. Consistently, full devices incorporating the SAM-modified interface showed performance improvements, with PCE increasing from (17.0 ± 0.7) % to (22.9 ± 0.4) % and $V_{OC}$ rising from (0.981 ± 0.019) V to (1.198 ± 0.004) V (**Fig. 2C**). trSPV measurements confirmed enhanced charge separation, consistent with the improved device performance. The amplitude of the SPV transients is negative due to the detection of surface electrons while holes are extracted towards NiO$_x$ (**Fig. 2D**). The SAM-modified interface exhibited faster charge separation within 100 ns, and a larger absolute trSPV amplitude than the bare NiO$_x$ interface, consistent with more efficient hole extraction. For the SAM-modified NiO$_x$/perovskite interface, PL intensity, QFLS, charge carrier lifetime, trSPV response, and device performance align, consistent with the conventional interpretation that a brighter PL reflects improved interface quality. Whether this interpretation holds for other interface classes will be explored next.

Interfaces limited by charge extraction are common in thin-film photovoltaics, often originating from energetic offsets or transport barriers at the CSC. In such cases, the number of charge carriers reaching the electrode is constrained, even when interface recombination is low. To explore how the extraction quality affects the PL response, we designed a heterojunction consisting of perovskite [FA$_{0.78}$Cs$_{0.22}$Pb(I$_{0.85}$Br$_{0.15}$)$_3$, FACs] interfaced with indene-C$_{60}$ monoadduct (ICMA).[41] By varying the ICMA layer thickness from 0.9 nm (thin) to 2.7 nm (mid) and 12.7 nm (thick; **Sec. S2.6.2**, Supporting Information), we controlled interfacial charge separation. This separation occurs when electrons and holes are extracted at unequal rates, leading to a spatial gradient in QFLS under open-circuit conditions.[42] Thicker ICMA layers permit a larger potential drop across the CSC, enhancing band bending.[27] These gradients in QFLS and band bending directly affect the observed PL and trSPV response. Since the chemical nature of the interface remains unchanged across all thicknesses, observed trends can be attributed to charge separation rather than to interface defects. We observed clear PL quenching with increasing ICMA thickness (**Fig. 2E**, **Fig. S1D**), accompanied by a reduction in QFLS from 1.19 eV for the thinnest ICMA to 1.10 eV for the thickest ICMA (**Fig. S1E**). The trPL decays became faster (**Fig. 2F**), and differential charge carrier lifetimes decreased accordingly (**Fig. S1F**). Consistent with decreasing QFLS for thicker ICMA, the $V_{OC}$ of full devices decreases from (1.153 ± 0.002) V to (1.125 ± 0.005) V (**Fig. S3F**). However, this $V_{OC}$ loss is overcompensated by an increase in fill factor (FF) from (65.2 ± 1.7) % to (72.7 ± 0.5) %, resulting in a notable PCE improvement from (16.0 ± 0.4) % to (17.8 ± 0.2) % (**Fig. 2G**). To understand the apparent contradiction between PL quenching and improved performance, we measured trSPV. Thicker ICMA layers exhibit more efficient charge separation, as indicated by a steeper trSPV rise and larger negative amplitudes within 20 ns after excitation (**Fig. 2H**). The negative sign of the trSPV signal results from the detection of electrons extracted toward the trSPV probe at the ICMA-exposed surface, while holes remain in the perovskite. Since ICMA thickness was the only varied parameter, enhanced charge extraction accompanies PL quenching. Despite the quenched PL, device performance improves, a trend that conventional PL analysis might have misinterpreted as poorer interface quality.

Finally, we examined a heterojunction where both interface recombination and charge extraction contribute simultaneously, represented by (doped) TiO$_2$/perovskite (CsPbI$_3$). Doping strategies are frequently used to adjust the work function (WF) of CSCs.[42-45] Doping can influence interface recombination by modifying the surface chemistry of the CSC, and affect charge extraction by shifting the energetic alignment at the interface. Therefore, we chose this interface to examine how combined changes in recombination and extraction influence PL and trSPV trends, and whether PL alone provides reliable guidance. In ssPL, we observed quenching upon doping $TiO_2$ (**Fig. 2I**, **Fig. S1G**), along with a slight decrease in QFLS from 1.12 eV to 1.11 eV (**Fig. S1H**), which could be interpreted as increased interface losses. However, trPL showed unexpected results: Both samples exhibited similar initial decays, but after 15 ns, the doped sample decayed more slowly (**Fig. 2J**), suggesting fewer interface losses. Solar cell device measurements supported this interpretation. Solar cells using doped $TiO_2$ achieved a higher PCE of $(17.4 \pm 0.4)$ % compared to $(16.4 \pm 0.5)$ % for $TiO_2$, primarily due to an increased FF, from $(78.8 \pm 1.9)$ % to $(80.8 \pm 1.8)$ % (**Fig. 2K**). To assess whether charge extraction also improved, we measured trSPV. The doped CSC exhibited a steeper initial slope and a higher trSPV amplitude than the reference CSC (**Fig. 2L**), indicating enhanced charge extraction. The observed trends in PL, trPL, trSPV, and device performance suggest that the doped system benefits from both reduced interface losses and enhanced extraction, combining characteristics of defect- and extraction-limited systems. However, the slight PL quenching would have misleadingly implied inferior performance when interpreted in isolation. In this class of interfaces, PL alone is insufficient to predict performance improvements. trSPV was essential to correctly identify the origin of the improvement.

Across the studied interfaces, we observed PL quenching even when device performance improved, a trend most pronounced in extraction-limited interfaces. This challenges the common assumption that PL intensity reliably reflects interface quality. PL reflected interface quality in defect-limited cases, but the correlation broke down as charge extraction improved. To resolve the ambiguous relationship between PL and device performance, we complemented PL with trSPV, which provides experimental access to charge extraction dynamics. trSPV clarified discrepancies between PL trends and device performance and demonstrated that PL quenching originates from efficient charge extraction. However, the detailed physical origin behind PL quenching remained unclear. A reduction in emitted photons may result from increased interfacial carrier densities, potentially due to transport limitations in the CSC,[46] or Coulomb attraction between electrons in the CSC and holes in the perovskite, which can increase local recombination without an increase in trap density.[28] To test these hypotheses, we employed drift-diffusion (DD) modeling, designed to reproduce the PL, trSPV and device behavior observed across the three classes of interfaces.

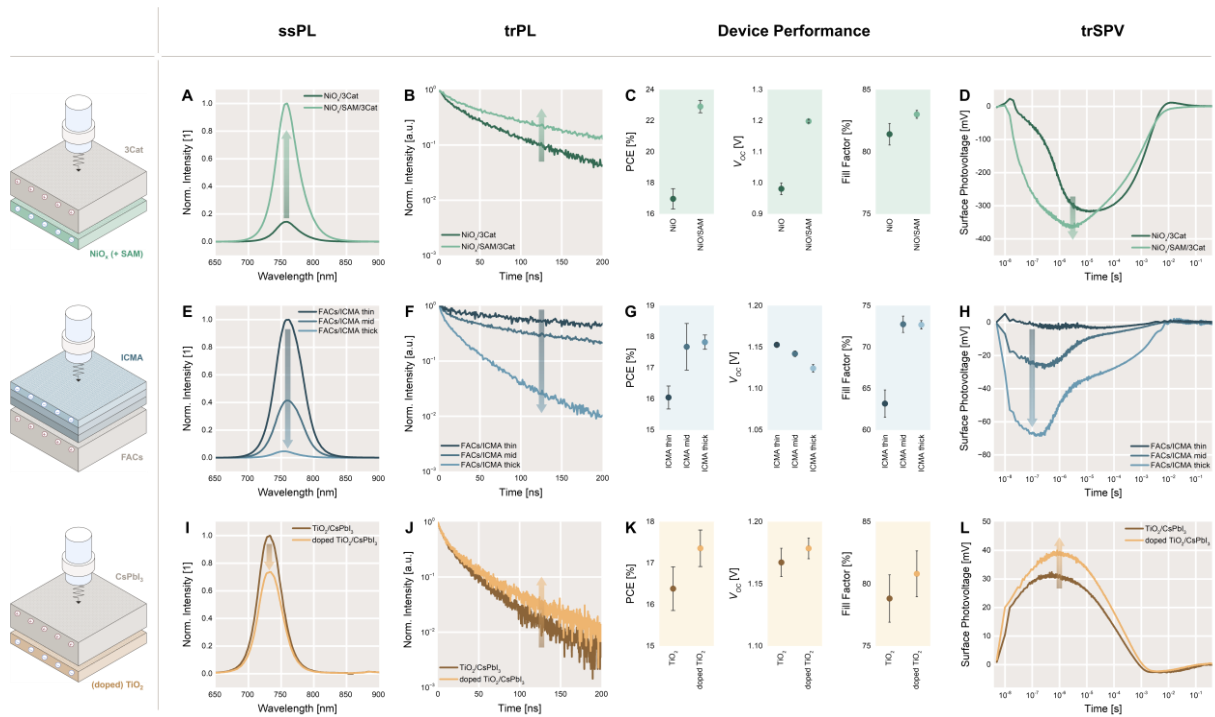

**Fig. 2. PL enhancement and quenching at perovskite/CSC interfaces showing improved charge extraction and device performance.** **(A)** ssPL, **(B)** trPL, **(C)** device performance metrics, and **(D)** trSPV of triple cation perovskite (3Cat) on $NiO_x$, bare and modified with a self-assembled monolayer (SAM). **(E)** ssPL, **(F)** trPL, **(G)** device performance metrics, and **(H)** trSPV of formamidinium cesium (FACs) mixed cation perovskite interfaced with ICMA layers of varying thickness: 0.9 nm (thin), 2.7 nm (mid), and 12.7 nm (thick). **(I)** ssPL, **(J)** trPL, **(K)** device performance metrics, and **(L)** trSPV of $CsPbI_3$ on (doped) $TiO_2$. Arrows indicate transitions from the control sample (darker curve) to the modified sample with enhanced charge extraction and performance (brighter curve).

Using drift-diffusion (DD) simulations, we built a digital replica of the semiconductor/electron-selective contact (ESC) heterojunction to investigate the mechanisms behind extraction-induced PL quenching. These simulations allow systematic investigation of Coulomb attraction, interfacial and bulk recombination, and energetic alignment, which control the photogenerated carrier redistribution and, in turn, PL quenching (**Fig. 1B**).[47,48] Unlike simpler rate-equation models,[34] the DD approach self-consistently solves Poisson's equation (**Eq. 1**) and the carrier continuity equations (**Eq. 2**, for electrons), capturing the spatial redistribution of charge carriers under illumination.[27] In **Eq. 1**, $E$ is the electric field, $x$ is the position, $q$ is the elementary charge, $\varepsilon$ is the material dielectric permittivity, $p$ and $n$ are the electron- and hole-concentrations, respectively, and $N_A^-$ and $N_D^+$ are the ionized acceptor and donor trap density, respectively. In **Eq. 2**, $t$ is the time, $\mu_e$ is the electron mobility, $D_e$ is the electron diffusion coefficient, $G_e$ is the generation rate, and $R_e$ is the recombination rate of electrons. $R_e$ includes bulk and interface Shockley-Read-Hall (SRH) terms $R_{\text{bulk,SRH}} = \frac{2(pn-n_i^2)}{\tau_{\text{bulk}}(n+p+2n_i)}$ and $R_{\text{interf,SRH}} = \frac{pn-n_i^2}{d_{\text{interf}}\left[\frac{(n+n_i)}{v_{\text{interf,h}}} + \frac{(p+n_i)}{v_{\text{interf,e}}}\right]}$ along with radiative and Auger contributions (**Eq. S11**), with $n_i$ being the intrinsic carrier concentration, $\tau_{\text{bulk}}$ being the SRH carrier lifetime under high-injection conditions, $d_{\text{interf}}$ being the interface thickness and $v_{\text{interf,e}}$ and $v_{\text{interf,h}}$ being the electron and hole interface recombination velocities, enabling spatially resolved modeling of radiative and non-radiative losses. Such a profound level of modeling is essential to understand how extraction and recombination processes both in the bulk and at interfaces determine the ssPL, trPL and trSPV signals. In our ESC/semiconductor model system with parameters of the $TiO_2/CsPbI_3$ heterojunction (**Fig. 2A**), we varied two central parameters: the interface hole recombination velocity ($v_{\text{interf,h}}$), to simulate interface passivation, and the conduction band offset ($\Delta\chi$), to simulate variations in charge extraction efficiency. In hole-selective heterojunctions, this approach is equivalent to varying the interface electron recombination velocity and valence band offset.

(**Eq. 1**) $$\frac{\partial E}{\partial x} = \frac{q}{\varepsilon}[p(x) - n(x) - N_A^- + N_D^+]$$

(**Eq. 2**) $$\frac{\partial n}{\partial t} = \mu_e E \frac{\partial n}{\partial x} + \mu_e n \frac{\partial E}{\partial x} + \frac{\partial}{\partial x}\left(D_e \frac{\partial n}{\partial x}\right) + G_e - R_e$$

We first tested the **effect of interface passivation** on the ssPL response by lowering $v_{\text{interf,h}}$ while keeping $\Delta\chi$ at constant 0.12 eV (**Sec. S3.4**). When interface recombination is fully suppressed, PL quenching does not occur, even though charges accumulate near the interface due to Coulomb attraction. This result confirms that PL quenching requires active non-radiative channels at the interface and does not result from charge accumulation alone. To show the **effect of charge extraction** on PL intensity, we varied $\Delta\chi$ (**Fig. 3A**) while keeping $v_{\text{interf,h}}$ constant. We set the characteristic time of bulk non-radiative recombination ($\tau_{\text{bulk}}$) to be 1.1 µs, while the characteristic time of interface non-radiative recombination ($\tau_{\text{interf}}$) was only 5.9 ns, so that $\tau_{\text{interf}} \ll \tau_{\text{bulk}}$. Reducing $\Delta\chi$ from 0.22 eV to 0.12 eV lowered the barrier for electron extraction. Under steady-state conditions, the bulk electron density decreased (**Fig. 3B**, left panel), and $R_{\text{NR}}$ in the perovskite bulk decreased accordingly (**Fig. S5H**). However, the extraction of electrons creates a local charge imbalance, drawing holes toward the interface via Coulomb attraction (**Fig. 3B**, right panel). Consequently, $R_{\text{NR}}$ at the interface increased (**Fig. 3C**) as accumulated holes continuously recombine with electrons re-injected from the selective contact. In the transient regime, larger trSPV amplitude (**Fig. 3D**) and accelerated trPL decay (**Fig. 3E**) are directly attributed to more efficient electron extraction occurring prior to recombination. As charge extraction improved with decreasing $\Delta\chi$, both the extracted electron density and the integrated $R_{\text{NR}}$ along the semiconductor thickness increased (**Fig. S12B**), resulting in PL quenching at steady-state conditions (**Fig. 3F**). This physical mechanism is illustrated in **Fig. 3G** where, at continuous generation and extraction, holes and electrons accumulate locally at both sides of the interface. Electrons re-injected from the selective contact rapidly recombine non-radiatively with accumulated holes, effectively quenching the PL. The condition $\tau_{\text{interf}} \ll \tau_{\text{bulk}}$ reproduces the extraction-limited behavior observed experimentally. Simulated *J-V* curves confirmed a reduction in extraction losses and an improvement in FF with decreasing $\Delta\chi$ due to lower series resistance (**Fig. S5E**), in line with experimental data (**Fig. S3G**). These results demonstrate that PL quenching under enhanced extraction is not caused by additional defects, but by Coulomb attraction that favors fast interface recombination, independent of the type of selective contact. Even with non-doped $TiO_2$, PL quenching occurs when the extraction barrier is lowered (**Fig. S10**).

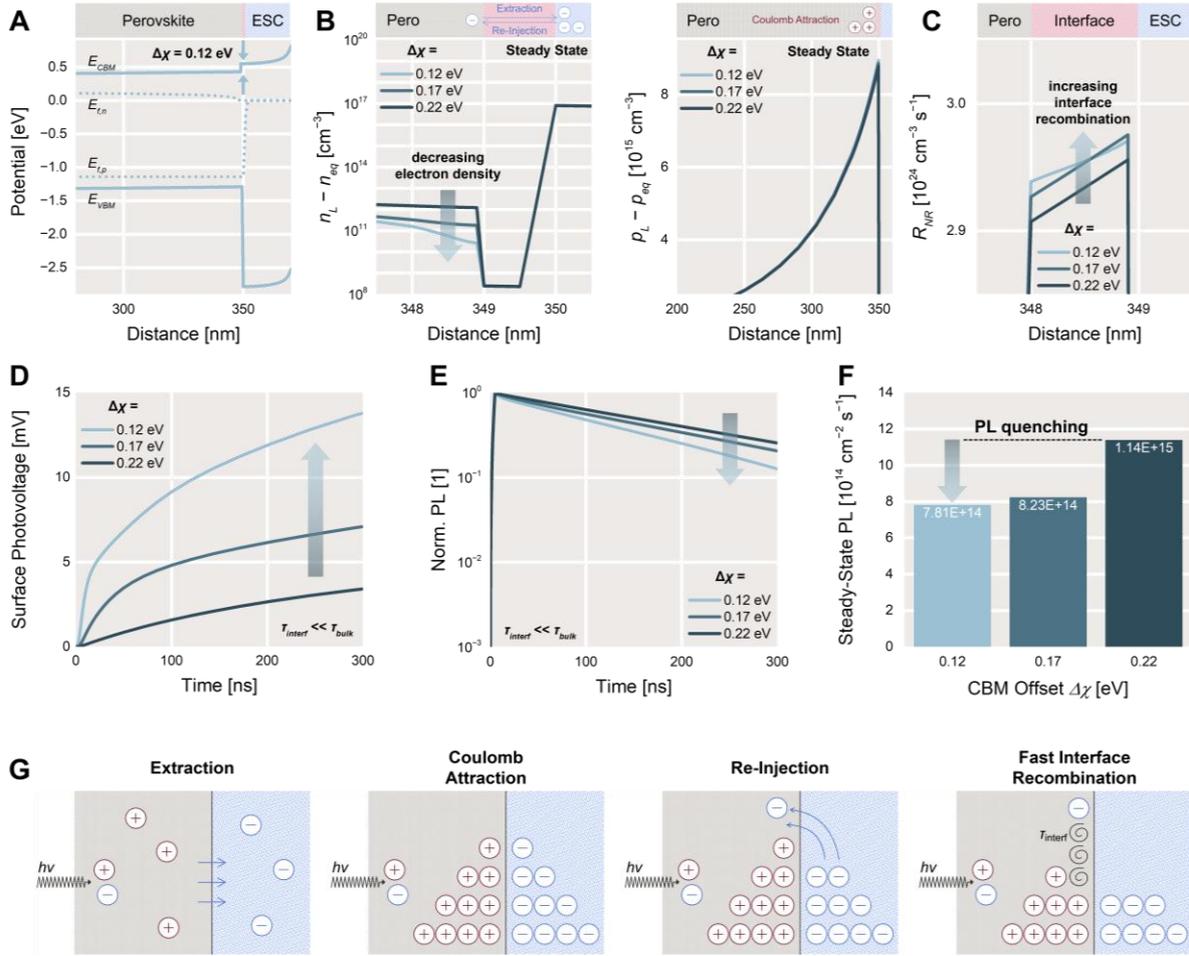

**Fig. 3. Digital replica reveals extraction-induced PL quenching at semiconductor interfaces. (A)** Out-of-equilibrium band diagram of the perovskite/ESC heterojunction for a conduction band offset $\Delta\chi$ = 0.12 eV. Reducing $\Delta\chi$ increases interfacial charge separation. **(B)** Steady-state charge carrier profiles: Lower $\Delta\chi$ depletes electron density in the bulk (left) and increases hole density at the interface due to Coulomb attraction to extracted electrons (right). **(C)** Spatially resolved $R_{NR}$: As extraction improves, $R_{NR}$ shifts from the bulk to the interface. **(D)** Simulated trSPV signals show enhanced extraction with decreasing $\Delta\chi$ ($\tau_{interf} \ll \tau_{bulk}$). **(E)** Simulated trPL transients exhibit faster decay at improved extraction. **(F)** Simulated ssPL intensity decreases with enhanced extraction, despite unchanged defect density. **(G)** Schematic of the extraction-induced PL quenching mechanism: Extracted electrons (left) induce hole accumulation (middle left); re-injected electrons recombine non-radiatively at the interface (middle right), reducing radiative emission (right).

To test the **limits of PL quenching**, we varied $\Delta\chi$ but set $\tau_{bulk}$ to 100 ns and disabled interface recombination ($\tau_{interf} \rightarrow +\infty$). These conditions mimic a system with high-quality interfaces and lower bulk quality ($\tau_{interf} \gg \tau_{bulk}$). In contrast to the extraction-limited system with strong interface recombination, we observed an increase in PL intensity under steady-state conditions (**Fig. 4A**). Without interface recombination, PL enhancement results from increased carrier separation in the bulk perovskite, which reduces non-radiative recombination, yielding a ~10% net increase in radiative recombination across the layer (**Fig. S6H**). In the transient regime, enhanced charge extraction increased the trSPV amplitude (**Fig. 4B**) and accelerated the trPL decay (**Fig. 4C**). Coulomb attraction also occurs in the transient regime, as demonstrated by the charge carrier profiles 40 ns after the excitation pulse (**Fig. 4D**, left panel for electrons, right panel for holes). The observation that trPL decay accelerates with increasing trSPV amplitude regardless of the magnitude of interface recombination confirms that early-time trPL decay is governed solely by charge extraction. In contrast, PL quenching at steady-state originates from increased interface recombination induced by charge extraction and Coulomb attraction.

Our results identify a continuous shift from PL quenching ($\tau_{interf} \ll \tau_{bulk}$) to PL enhancement ($\tau_{interf} \gg \tau_{bulk}$, **Fig. 4E**), explaining the differing PL responses across the heterojunction types. The shift passes through a **balance point**, where PL enhancement due to interface passivation and PL quenching due to charge extraction compensate each other, thus marking the $\tau_{interf}/\tau_{bulk}$ ratio where ssPL intensity remains unchanged. Such competing effects on PL are present in the third class of interfaces, represented by $TiO_2/CsPbI_3$, where $TiO_2$ doping simultaneously lowers the interface defect density and improves energetic alignment. Variation of both parameters in the simulation (**Tab. S10-11**) qualitatively reproduces the experimentally observed PL quenching, but at the same time slower trPL decay, increased PCE, and enhanced trSPV (**Fig. S13A-D**), demonstrating the accuracy of our model even in such complex scenarios.

Our experiments and simulations demonstrate how both charge extraction and interface recombination affect the PL response in semiconductor/CSC heterojunctions. Efficient extraction increases the interfacial charge carrier density via Coulomb attraction, which enhances interface recombination and, consequently, quenches the PL signal, even as device performance improves. Although increased interface recombination can reduce $V_{OC}$, this loss is consistently overcompensated by a pronounced increase in FF, making FF the more reliable indicator of device performance. trSPV provides direct insight into charge extraction dynamics and FF, complementing PL's sensitivity to non-radiative losses that affect $V_{OC}$. Our findings further explain why interface engineering strategies such as doping, dipole insertion, or band alignment tuning can reduce PL while simultaneously improving device performance. Interface modification strategies previously dismissed based on PL alone may need to be reevaluated. The combination of PL and trSPV enables reliable, non-invasive characterization of semiconductor/CSC interfaces, and is compatible with high-throughput, contactless screening workflows central to automated device optimization.

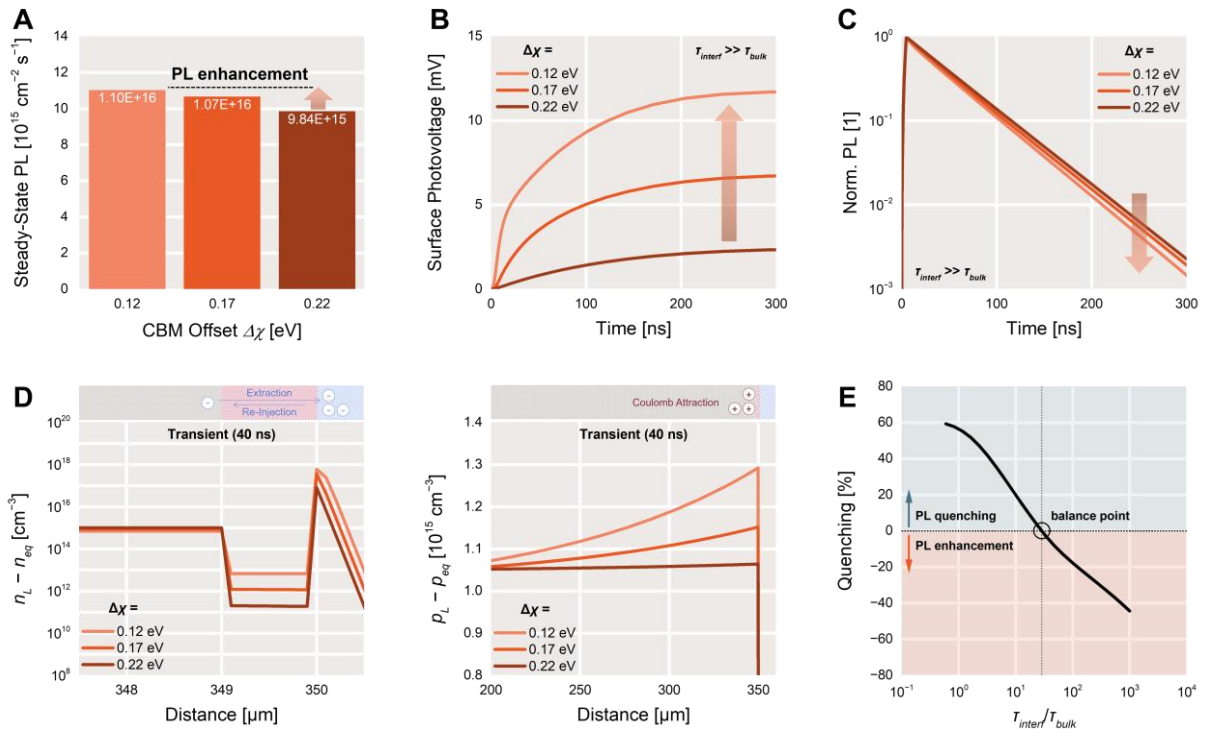

**Fig. 4. The balance between interface and bulk recombination defines the PL response. (A)** When interface recombination is slow relative to the bulk ($\tau_{interf} \gg \tau_{bulk}$), ssPL intensity increases with improved extraction. **(B, C)** Simulated trSPV and trPL transients exhibit the same trends as extraction-limited interfaces, confirming that early-time trPL decay is governed by charge separation dynamics. **(D)** Simulated charge carrier profiles shortly after excitation show that Coulomb attraction during extraction (left) induces hole accumulation at the interface (right). **(E)** The overall ssPL response transitions from quenching to enhancement as the ratio $\tau_{interf}/\tau_{bulk}$ increases, establishing this ratio as a core parameter for interpreting PL signals of semiconductor/CSC heterojunctions.

For our heterojunction system, we observed a $\tau_{interf}/\tau_{bulk}$-dependent transition from PL quenching to enhancement. To generalize the findings to highest-efficiency devices, we simulated the simple $TiO_2/CsPbI_3$ system of **Sec. S3.4.2** with an extended range for $\tau_{bulk}$ (1 ns to 10 µs) and $\tau_{interf}$ (10 ps to 100 ms). The PCE and PL response were calculated at 1 sun intensity for an extraction barrier $\Delta\chi$ of +0.15 eV (low extraction scenario), and a facilitated electron extraction ($\Delta\chi$ of −0.15 eV, high extraction scenario). Increasing $\tau_{bulk}$ and $\tau_{interf}$, whether individually or simultaneously, leads to continuous increase of the PCE both in the low- and in the high-extraction scenarios (**Fig. S16A-B**). The maxima of the PCE are at 20.9% for the barrier-type, and 26.2% for the cliff-type heterojunction. Reducing the extraction barrier generally improves the PCE of the solar cell (**Fig. 5A**).

To relate improvements in PCE to the system's PL response, we compared the high- and low-extraction scenarios and mapped whether the PL was quenched (positive values) or enhanced (negative values; **Fig. 5B**). The resulting two-dimensional map shows four distinct regions separated by a diagonal axis. Two "wings" extend symmetrically above and below this diagonal, representing the two opposite PL responses. The lower (azure) wing corresponds to PL quenching due to bulk lifetimes being higher than interface lifetimes ($\tau_{interf} \ll \tau_{bulk}$), while the upper (orange) wing indicates PL enhancement due to interface lifetimes being higher than bulk lifetimes ($\tau_{interf} \gg \tau_{bulk}$). This behavior is consistent with the one-dimensional cases discussed previously. A third region is defined by one of both lifetimes being very short ($\tau_{bulk} < 10^{-8}$ s; $\tau_{interf} < 10^{-10}$ s), where the perovskite is in the low-injection regime, and improved extraction efficiency causes PL quenching. This scenario is likely relevant for present tin-based perovskites. Finally, the upper right corner of the map corresponds to highest bulk and interface qualities, where the perovskite approaches the radiative limit, and charge extraction no longer influences the PL intensity.

Although the PCE improves due to improved extraction, both PL quenching and enhancement can occur, as shown by the corresponding regions of **Fig. 5B** highlighted in **Fig. 5A**. **Fig. 5C**, obtained by increasing $\tau_{interf}$ while maintaining $\tau_{bulk} = 10\ \tau_{interf}$ confirms that PL quenching does not correlate with the PCE improvement achieved by reducing extraction losses. In contrast, an increase in trSPV amplitude serves as a reliable indicator of enhanced PCE, since the trSPV signal is independent of $\tau_{bulk}$ and $\tau_{interf}$. When extraction losses are reduced, both PCE and trSPV increase (**Fig. 5D**). This analysis demonstrates that the PL signal can only be interpreted reliably when the characteristic lifetimes $\tau_{bulk}$ and $\tau_{interf}$ are known.

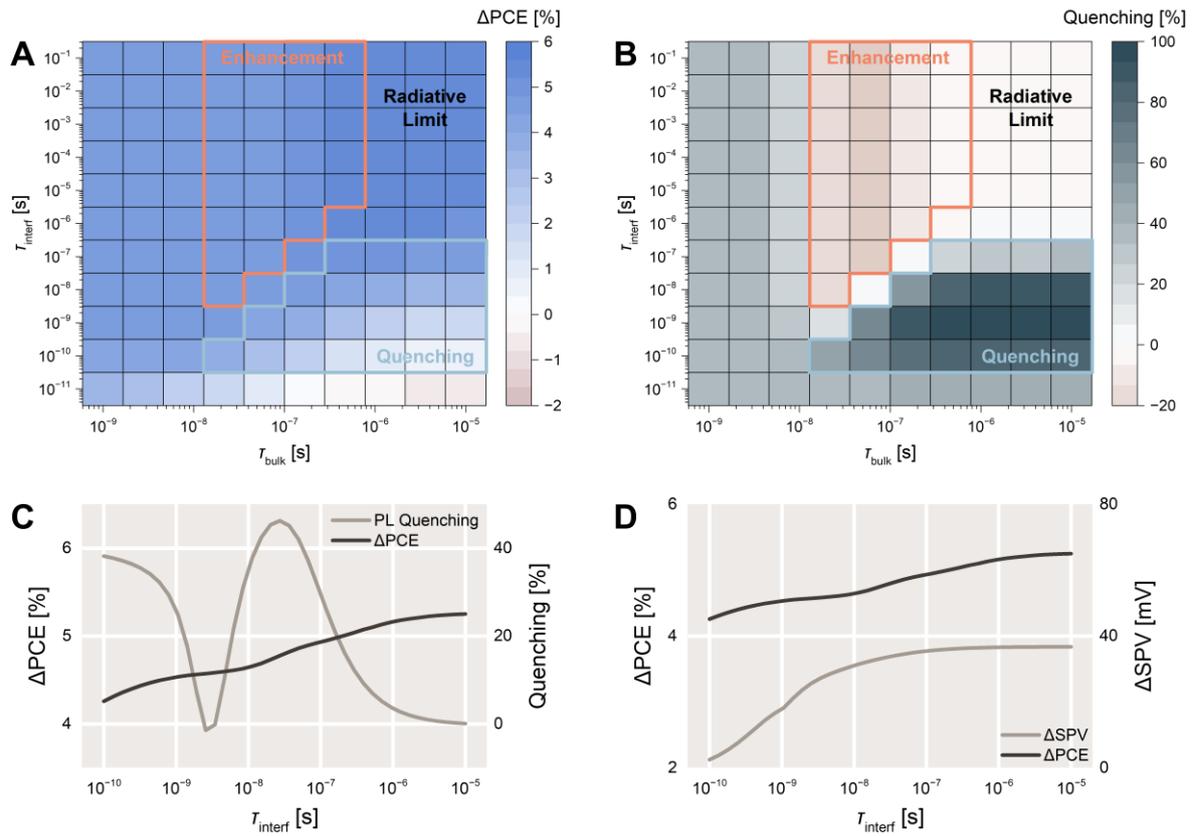

**Fig. 5. 2D maps as a function of the characteristic times $\tau_{bulk}$ and $\tau_{interf}$ of CsPbI$_3$ on TiO$_2$.** (**A**) Difference of PCE when the heterojunction is turned from a barrier ($\Delta\chi = +0.15$ eV) to a cliff ($\Delta\chi = -0.15$ eV), representing an improvement in charge extraction. (**B**) Corresponding change in PL intensity under 1-sun illumination, with negative values (orange) representing PL enhancement, and positive values (azure) representing PL quenching. (**C**) ΔPCE and PL quenching for increasing $\tau_{bulk}$ and $\tau_{interf}$, while keeping $\tau_{bulk} = 10\,\tau_{interf}$. (**D**) ΔPCE and ΔSPV amplitude, both calculated comparing the low- and high-extraction scenarios, for increasing $\tau_{bulk}$ and $\tau_{interf}$, while keeping $\tau_{bulk} = 10\,\tau_{interf}$.

To summarize our results, we translated the insights obtained from our digital replica into a decision tree (**Fig. 6**) that allows to classify interface modifications using only PL and trSPV measurements. The tree compares semiconductor/CSC interface modifications based on ssPL intensity, trPL decay, and trSPV amplitude relative to a reference. The first branch divides the tree into cases where ssPL is enhanced (A) or quenched (B). Among the three classes of heterojunctions studied in detail, PL quenching was observed for the FACs/ICMA (blue) and $TiO_2$/$CsPbI_3$ (yellow) interfaces, while PL enhancement occurred for the NiO/SAM/3Cat interface (green). Further branches assess whether trPL decay becomes slower (A) or faster (B) and whether trSPV amplitude increases (A) or decreases (B). This yields $2^3$ terminal cases, labeled **AAA** through **BBB**, which group into four physically distinct interface scenarios: increased or decreased interface defect density, affecting $v_{interf}$ (green scenarios); impaired or improved charge extraction due to energetic barriers ($\Delta\chi$) when $\tau_{interf} \ll \tau_{bulk}$ (blue scenarios); impaired or improved extraction when $\tau_{interf} \gg \tau_{bulk}$ (red scenarios); and mixed cases where $v_{interf}$ and $\Delta\chi$ vary simultaneously (yellow scenarios).

To test the generality of the decision tree, we classified more than 80 additional semiconductor/CSC heterojunctions, all characterized in our laboratory using the same protocols for ssPL, trPL, and trSPV under open-circuit conditions. Representative examples covering a wide range of interface types are included in the decision tree.[38,49-53] These examples validate the decision tree as a broadly applicable tool for classifying interfaces according to their optoelectronic response, revealing underlying physical mechanisms, and guiding the rational design of new semiconductor/CSC heterojunctions.

When combined with PL, trSPV enables reliable interface characterization by directly probing charge separation dynamics. The decision tree highlights our central finding that trSPV is essential for distinguishing efficient from inefficient charge extraction, and passivation effects from extraction barriers. Together, PL and trSPV measurements resolve the long-standing misconception that associates PL quenching with increased interface defect density, and enable systematic, contactless classification of heterojunctions in optoelectronic devices. The approach is compatible with high-throughput screening and supports routine laboratory workflows.[18]

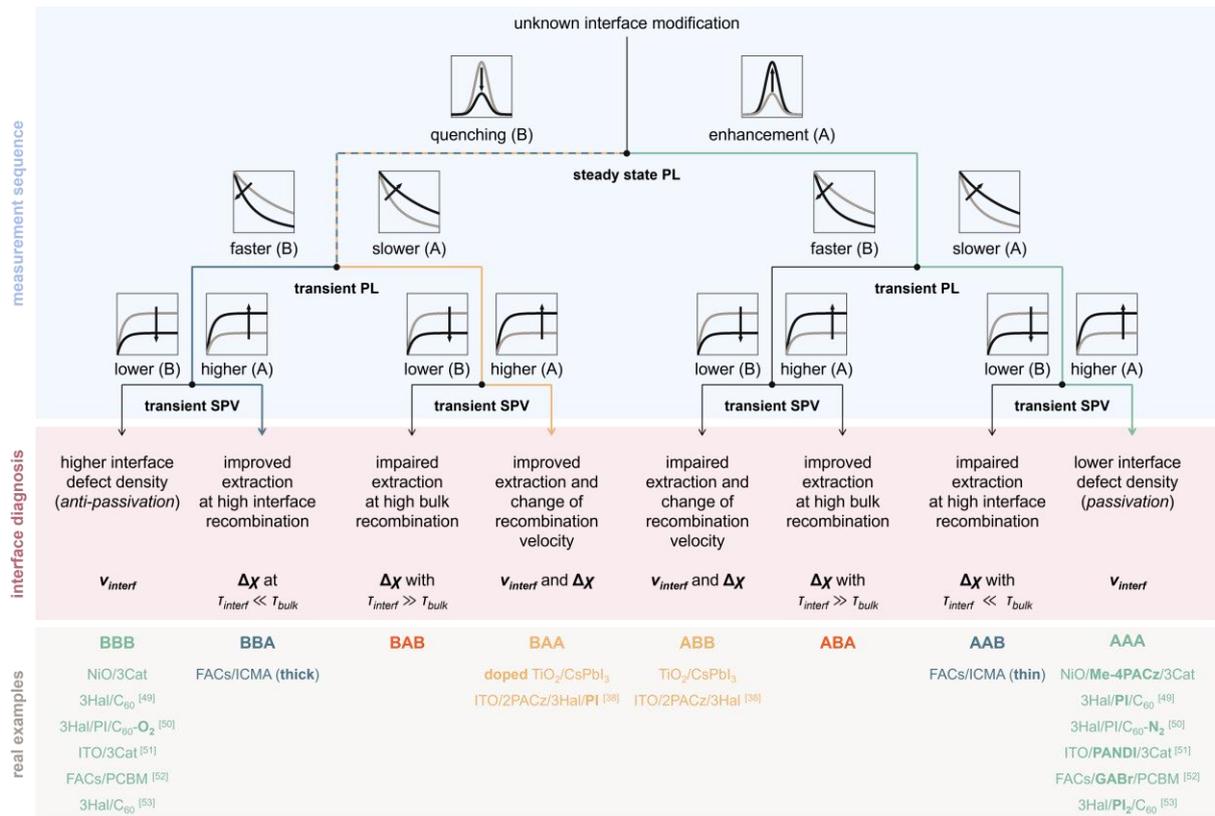

**Fig. 6. Classification of heterojunction behavior based on PL and trSPV measurements.** The decision tree distinguishes semiconductor/CSC interface modifications by comparing ssPL intensity, trPL decay, and trSPV amplitude relative to a reference. Each of the three branch points corresponds to one measurement criterion, resulting in eight cases grouped into four contrasting interface scenarios: Variation in interface defect density ($v_{interf}$, cases AAA and BBB), extraction barriers under $\tau_{interf} \ll \tau_{bulk}$ ($\Delta\chi$, cases AAB and BBA), extraction barriers under $\tau_{interf} \gg \tau_{bulk}$ ($\Delta\chi$, cases ABA and BAB), and mixed cases with simultaneous changes in $v_{interf}$ and $\Delta\chi$ (cases ABB and BAA). The three heterojunctions studied in detail, NiO/SAM/3Cat (green), FACs/ICMA (blue), and TiO$_2$/CsPbI$_3$ (yellow), are highlighted. Additional representative interfaces, all characterized in our laboratory under identical protocols, are classified according to their optoelectronic response, with respective modifications marked in bold.[38,49-53]

# Conclusion

Despite rapid progress in high-throughput materials discovery, current screening workflows often rely on photoluminescence (PL) as a proxy for material quality, while overlooking the role of interfaces in determining device performance. We show that PL measurements alone cannot distinguish whether PL quenching arises from enhanced non-radiative recombination or from efficient charge extraction, limiting their interpretability in heterojunctions and half-devices. We demonstrated that PL quenching, frequently attributed to interface defects, may instead result from efficient charge extraction, particularly in high-performance perovskite/charge-selective contact heterojunctions. To resolve this ambiguity, we complemented PL with transient surface photovoltage measurements, which directly probe charge extraction independently of radiative processes. Applied across three interface classes, this combined approach allowed extraction and recombination dynamics to be disentangled. To gain fundamental insight into the observed phenomenon, we developed a drift-diffusion-based digital replica, which reveals that fast extraction induces uncompensated charge in the CSC. This extracted charge triggers Coulomb attraction, drawing opposite charge carriers to the interface and increasing interface non-radiative recombination, causing PL quenching even as device performance improves.

These findings demonstrate that PL alone is insufficient for reliable interface assessment and leads to misinterpretation during materials screening. In contrast, trSPV provides direct, non-invasive insight into charge extraction and interfacial behavior, offering predictive value for performance metrics such as fill factor and open-circuit voltage. We propose a decision tree that classifies $2^3$ distinct interface scenarios using PL and trSPV trends. The decision tree represents a practical tool for evaluating semiconductor heterojunctions without requiring full device fabrication. By enabling interface assessment at the level of partial devices, our approach addresses a critical bottleneck in heterojunction quality prediction. This supports rational interface engineering and design across photovoltaics, photodetectors, and LEDs. The methodology is compatible with automated platforms and bridges the gap from accelerated materials discovery to accelerated device discovery.

## Acknowledgments

The authors want to thank the technicians Carola Ferber, Markus Johannes Beckedahl, Michel Choyne, Hagen Heinz, and Thomas Lußky for making the HySPRINT laboratory run smoothly. We gratefully acknowledge Dr. Thomas Dittrich for providing the HZB SPV laboratory facilities. We acknowledge HyPerCells – a joint graduate school of the University of Potsdam and HZB. This project has received funding from the German Federal Ministry of Education and Research (Bundesministerium für Bildung und Forschung, BMBF) under the NanoMatFutur Call, project number 03XP0625, COMET PV, and the European Union's Framework Program for Research and Innovation HORIZON EUROPE (2021-2027) under the Marie Skłodowska-Curie Action Postdoctoral Fellowships (European Fellowship) 101061809 HyPerGreen. T. W. Gries and A. Musiienko gratefully acknowledge the financial support of the German Federal Ministry of Education and Research (BMBF) within the project "Transatlantische Exzellenz-Allianz für PV Innovationen" (TEAM PV) with funding code 03SF0747. D. Regaldo acknowledges the EU Horizon Valhalla project for funding his research (grant agreement number 101082176). T. W. Gries thanks Lea Zimmermann for valuable discussions. The authors gratefully acknowledge Barbara Herschel (*Bureau Barbara & Partner*, Berlin; e-mail: bureau.barbara@web.de) for the graphic design.

## Author Contributions

T. W. Gries and D. Regaldo conducted the research. T. W. Gries fabricated photovoltaic devices and samples containing $TiO_2$, and recorded and analyzed ssPL data, trSPV data, and *J-V* data. D. Regaldo developed the DD model with the assistance of A. Musiienko and J.-P. Kleider. D. Regaldo fitted the trSPV data in the system containing $TiO_2$. Y. Duan fabricated photovoltaic devices containing $NiO_x$, and recorded *J-V* data. F. Scheler fabricated photovoltaic devices containing ICMA, and recorded *J-V* data. M. Simmonds recorded and analyzed trPL data and provided discussion on the simulation and theoretical sections. V. Stacchini supported the recording of trSPV data and provided fruitful discussion. A. Petrozza provided guidance to D. Regaldo. A. Abate provided guidance to T. W. Gries. A. Musiienko and J.-P. Kleider provided fundamental discussion and guidance. T. W. Gries and D. Regaldo prepared and designed the manuscript revisited by J.-P. Kleider and A. Musiienko. A. Musiienko took the supervision of the project.

## Conflict of Interest

The authors do not declare conflict of interest.

## Data Availability

The data supporting this article have been included as part of the Supporting Information.